# CUSTOMER DATA CLUSTERING USING DATA MINING TECHNIQUE

Dr. Sankar Rajagopal

Enterprise DW/BI Consultant
Tata Consultancy Services, Newark, DE, USA
e-mail:rajagopal.sankar@tcs.com

## ABSTRACT:

*Classification and patterns extraction from customer data is very important for business support and decision making. Timely identification of newly emerging trends is very important in business process. Large companies are having huge volume of data but starving for knowledge. To overcome the organization current issue, the new breed of technique is required that has intelligence and capability to solve the knowledge scarcity and the technique is called Data mining. The objectives of this paper are to identify the high-profit, high-value and low-risk customers by one of the data mining technique - customer clustering. In the first phase, cleansing the data and developed the patterns via demographic clustering algorithm using IBM I-Miner. In the second phase, profiling the data, develop the clusters and identify the high-value low-risk customers. This cluster typically represents the 10-20 percent of customers which yields 80% of the revenue.*

## KEYWORDS:

*Data mining, customer clustering and I-Miner*

# 1. INTRODUCTION

For a successful business, identification of high-profit, low-risk customers, retaining those customers and bring the next level customers to above cluster is a key tasks for business owners and marketers. Traditionally, marketers must first identify customer cluster using a mathematical mode and then implement an efficient campaign plan to target profitable customers (1-4). This process confronts considerable problems. Most previous studies used various mathematical models to segment customers without considering the correlation between customer cluster and a campaign/loyalty programs. Moreover, due to advances in computing and information storage areas large companies are piling up vast volume of data and the traditional mathematical models are difficult predicts the segmentations and patterns. As a result, useful information is often overlooked, and the potential benefits of increased computational and data gathering capabilities are only partially realized. Only through data mining techniques, it is possible to extract useful pattern and association from the customer data (5). Data mining techniques like clustering and associations can be used to find meaningful patterns for future predictions (6,7).Customer clustering and segmentation are two of the most important techniques used in marketing and customer-relationship management. They use customer-purchase transaction data to track buying behavior and create strategic business initiatives. Businesses can use this data to divide customers into segments based on such "shareholder value" variables as current customer profitability, some measure of risk, a measure of the lifetime value of a customer, and retention probability. Creating customer segments based on such variables highlights obvious marketing opportunities.

The paper is organized as follows. Section 2 discuss about Background research. Section 3 briefly reviews data mining and different clustering techniques. The proposed architecture, experiments and results are discussed in the section 4. Section 5 concludes the paper and gives suggestions for future work.

## 2.  RESEARCH BAGROUND

In traditional markets, customer clustering / segmentation is one of the most significant methods used in studies of marketing. This study classifies existing customer cluster/segmentation methods into methodology-oriented and application-oriented approaches. Most methodology-driven studies used mathematical methodologies; e.g statistics, neural net, generic algorithm (GA) and Fuzzy set to identify the optimized segmented homogenous group (8-15).

In recent years, it has been recognized that the partitioned clustering technique is well suited for clustering a large dataset due to their relatively low computational requirements. Behavioral clustering and segmentation help derive strategic marketing initiatives by using the variables that determine customer shareholder value. By conducting demographic clustering and segmentation within the behavioral segments, we can define tactical marketing campaigns and select the appropriate marketing channel and advertising for the tactical campaign. It is then possible to target those customers most likely to exhibit the desired behavior by creating predictive models.

In this work demographic clustering algorithm is used to identify the customer clustering. In phase 1, the customer data is cleansed and developed patterns using various parameters and subsequently, in phase 2 profiled the data, developed the clusters and identified the high-value low risk customers. From the experimental results it showed that the proposed approach would generate more useful pattern from large data.

## 3. DATA MINING AND CLUSTERING METHODS

Data mining - also known as knowledge-discovery in databases (KDD) is process of extracting potentially useful information from raw data. A software engine can scan large amounts of data and automatically report interesting patterns without requiring human intervention. Other knowledge discovery technologies are Statistical Analysis, OLAP, Data Visualization, and Ad hoc queries. Unlike these technologies, data mining does not require a human to ask specific questions.

In general, Data mining has four major relationships. They are:

    (i)  Classes
    (ii) Clusters

(iii) Associations
(iv) Sequential patterns.

**(i)**    **Classes:** Stored data is used to locate data in predetermined groups. For example, a restaurant chain could mine customer purchase data to determine when customers visit and what they typically order. This information could be used to increase traffic by having daily specials.

**(ii)**   **Clusters:** Data items are grouped according to logical relationships or consumer preferences. For example, data can be mined to identify market segments or consumer affinities.

**(iii)**  **Associations:** Data can be mined to identify associations. The beer-diaper example is an example of associative mining.

**(iv)**   **Sequential patterns:** Data is mined to anticipate behavior patterns and trends. For example, an outdoor equipment retailer could predict the likelihood of a backpack being purchased based on a consumer's purchase of sleeping bags and hiking shoes.

## 3.1 Clustering Methods:

Clustering is a typical unsupervised learning technique for grouping similar data points. A clustering algorithm assigns a large number of data points to a smaller number of groups such that data points in the same group share the same properties while, in different groups, they are dissimilar. Clustering has many applications, including part family formation for group technology, image segmentation, information retrieval, web pages grouping, market segmentation, and scientific and engineering analysis (16).

Many clustering methods have been proposed and they can be broadly classified into four categories (17-24): partitioning methods, hierarchical methods, density-based methods and grid-based methods. Other clustering techniques that do not fir in these categories have been developed. They are fuzzy clustering, artificial neural networks and generic algorithms.

The following section deals about detailed study of the customer clustering. The data is the production information of our organization smart retail store.

## 3.2 Customer Clustering:

Customer clustering is the most important data mining methodologies used in marketing and customer relationship management (CRM). Customer clustering would use customer-purchase transaction data to track buying behavior and create strategic business initiatives.

Companies want to keep high-profit, high-value, and low-risk customers. This cluster typically represents the 10 to 20 percent of customers who create 50 to 80 percent of a company's profits. A company would not want to lose these customers, and the strategic initiative for the segment is obviously retention. A low-profit, high-value, and low-risk customer segment is also an attractive one, and the obvious goal here would be to increase profitability for this segment. Cross-selling (selling new products) and up-selling (selling more of what customers currently buy) to this segment are the marketing initiatives of choice.

### 3.3 Proposed Architecture:

The proposed approach is a two phased model. In first phase, collect the data from our organization retail smart store and then do the data cleansing. It involves removing the noise first, so the incomplete, missing and irrelevant data are removed and formatted according to the required format. In second phase, generate the clusters and profile the clusters to identify by best clusters. Fig.1 illustrates the whole process.

Fig: 1 Clustering Process

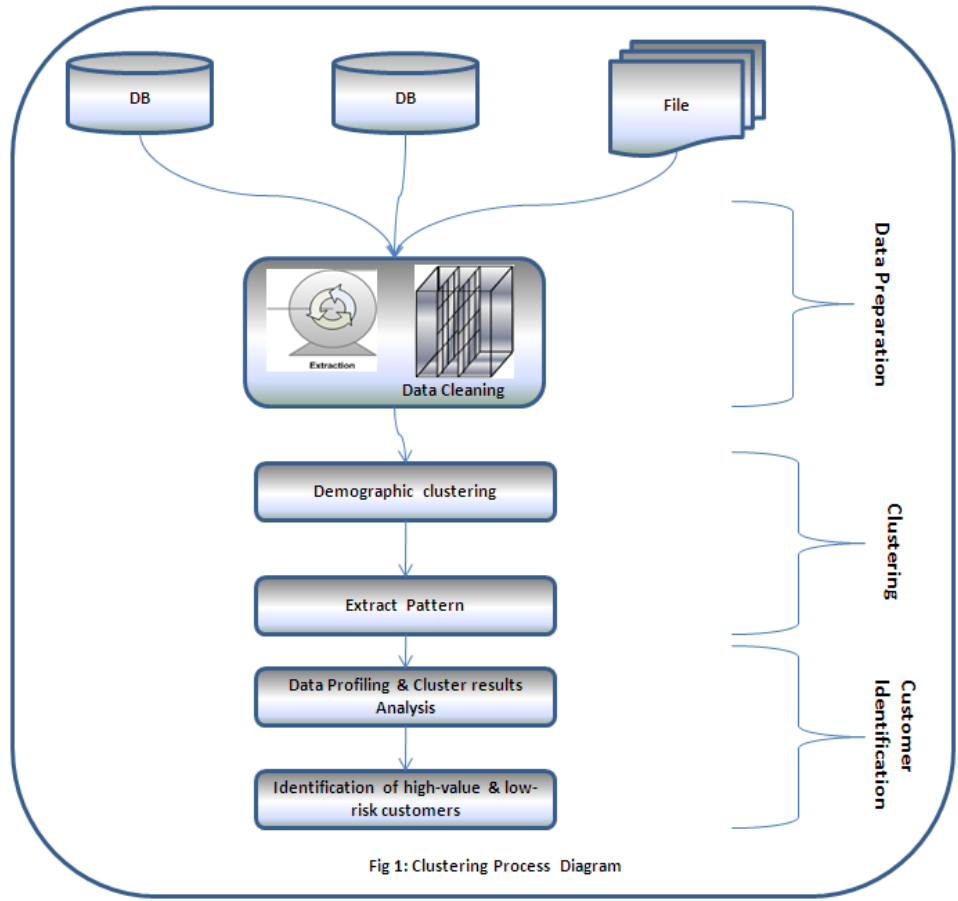

Fig 1: Clustering Process Diagram

## 4. EXPERIMENTS AND RESULTS:

For this study, the transaction of data of our organization retail smart store has been taken. Using these data, customers have been clustered using IBM Intelligent Miner tool. The first steps in the clustering process involve selecting the data set and the algorithm. There are two types of algorithms available in I-Miner process.

      i.   Demographic clustering process
     ii.   Neural clustering process.

In this exercise, the Demographic clustering process has been chosen, since it works best for the continuous data type. The data set has all the data types are continuous.

The next step in the process is to choose the basic run parameters for the process. The basic parameters available for demographic clustering include are:

- Maximum number of clusters
- Maximum number of passes through the data
- Accuracy
- Similarity threshold

For this assignment, the maximum number of clusters is 4, maximum number of passes through the data is 3 and the accuracy is 0.5 have been chosen.

The input parameters for the customer's clustering are:

1. Recency
2. Total customer profit
3. Total customer revenue
4. Top revenue Department

The top revenue department variable has been chosen as supplementary variable and rest of the variables have been chosen as active variables. The supplementary variables are used to profile the clusters and not to define them. The ability to add supplementary variables at the outset of clustering is a very useful feature of Intelligent Miner that allows quick and easy interpretation of clusters using data other than the input variables. The input and output field's width are defined and The input data used in mining is the production data of our organization retail smart store. (Synthesized based on the experience).

The data is first extracted from the oracle databases and flat files and converted into flat files. Subsequently, the IMiner process picks up the file and processed. The following figures show the snapshot of some of the IMiner process output screens. The entire output data set would have customer information appended to the end of the each record.

The below figures 2  show the results of the clustering run in Intelligent Miner's cluster visualizer, which was used by both demographic and neural clustering (clustering mode). Each of the horizontal strips represents a cluster (with its ID number on the left). The clusters are ordered from top to bottom in order of size, with the number to the penultimate left indicate ng the size of a cluster as a percentage of the universe.

Fig. 2: I-Miner Results

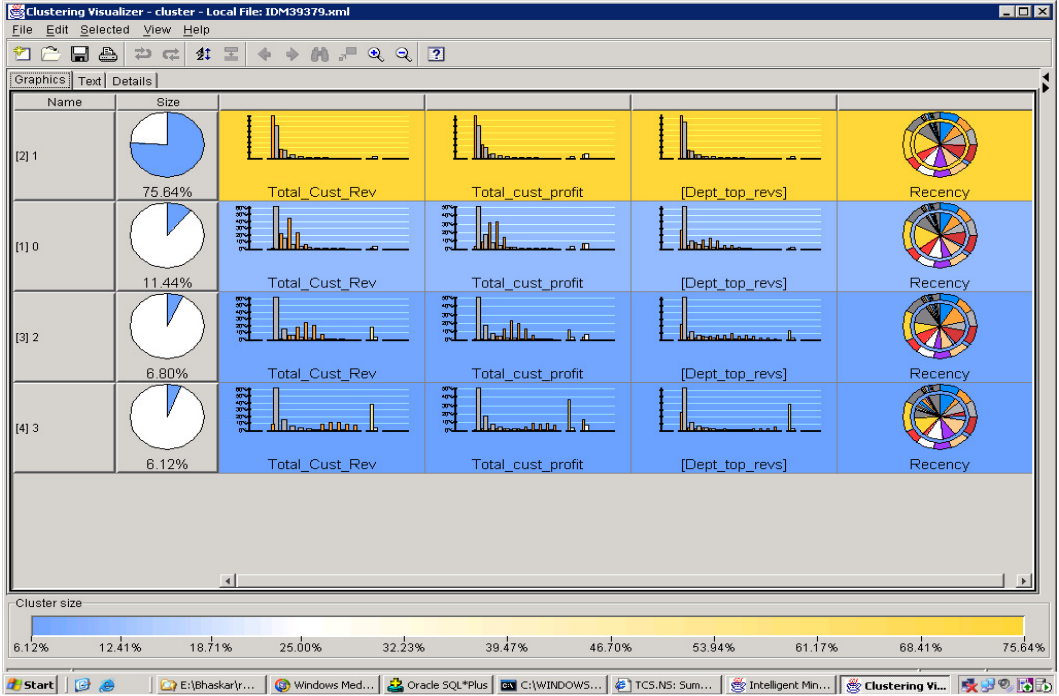

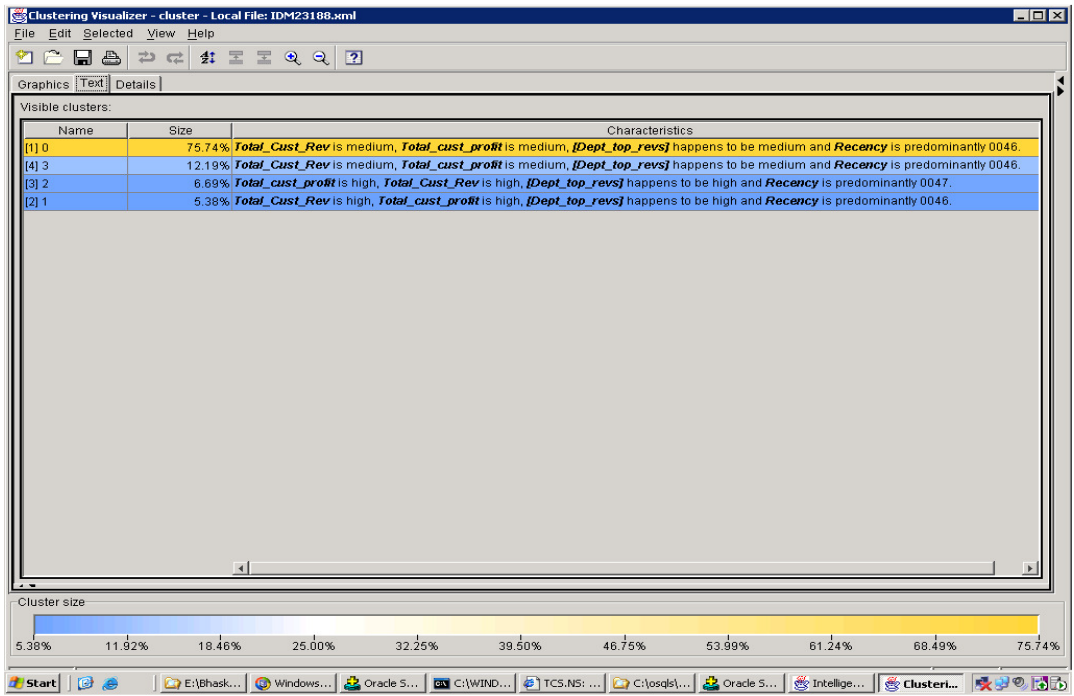

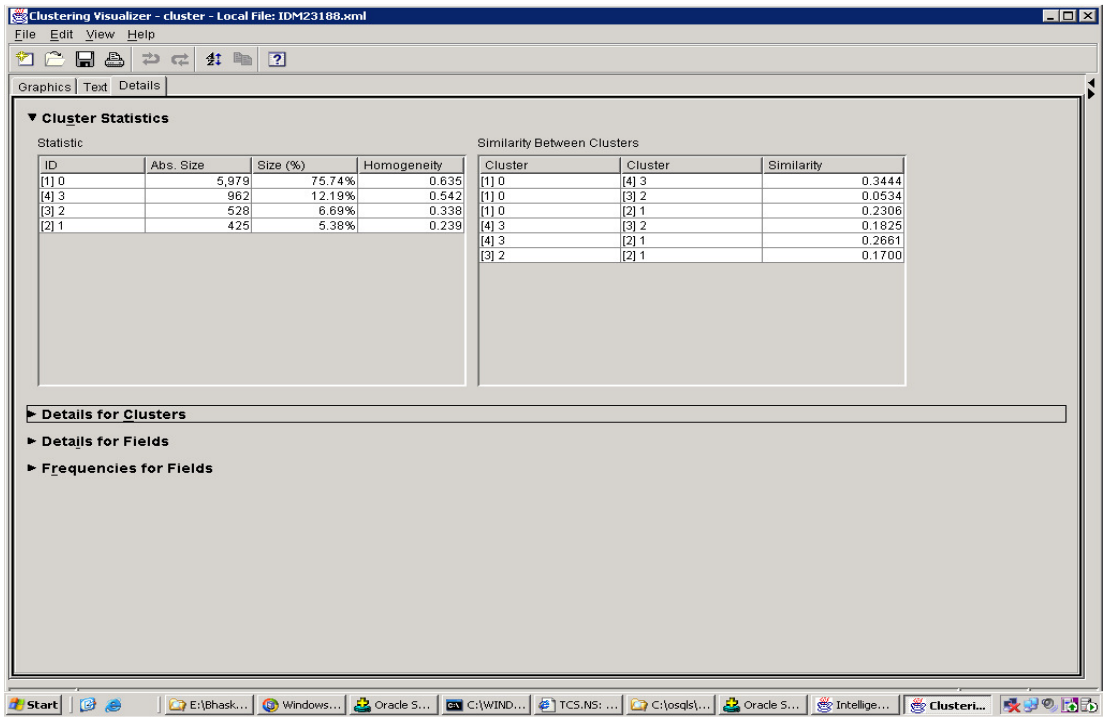

Variables are ordered from left to right in order of importance to the cluster, based on a chi-square test between variable and cluster ID. This is the default metric. Other metrics include entropy, Condorcet criterion, and database order.

The following figure shows the characteristics of the clusters along with cluster id (Name) and size of the each cluster. The variables used to define clusters are without brackets, while the supplementary variables appear within brackets. Numeric (integer), discrete numeric (small integer), binary, and continuous variables have their frequency distribution or histogram shown as a bar graph. The red bars in the foreground indicate the distribution of the variable within the current cluster. The gray solid bars in the background indicate the distribution of the variable in the whole universe. The more different the cluster distribution is from the average, the more interesting or distinct the cluster.

The output of the I-miner clustering would be generated as a fixed length flat file which contains the output fields along with cluster ids, condorcet values and the confidence values. Subsequently, the flat file would be loaded into ORACLE table for profiling using SQL loader.

## 4.1 Cluster Profiling

The next step in the clustering process is to profile the clusters by executing SQL queries. The purpose of profiling is to assess the potential business value of each cluster quantitatively by profiling the aggregate values of the shareholder value variables by cluster.

Below table 1 provides values of a profile of revenue, profit and total revenue in percentage. The table shows that cluster 1 is the most profitable cluster, representing about 35 percent of the revenue yet only 6 percent of the customers.

Table 1:  Clustering results with total revenue, customer count, total profit etc.

| CLUSTER RANK | CLUSTER ID | CUSTOMER COUNT | TOTAL PROFIT | TOTAL REVENUE | CUSTOMER COUNT(%) | TOTAL REV.(%) |
|---|---|---|---|---|---|---|
| 1 | 3 | 486 | 117256.59 | 563643.71 | 6.16 | 35.47 |
| 2 | 1 | 5977 | 85812.08 | 382009.43 | 75.72 | 24.04 |
| 3 | 2 | 529 | 82169.06 | 379789.03 | 6.70 | 23.90 |
| 4 | 0 | 902 | 51561.46 | 263656 | 11.43 | 16.59 |

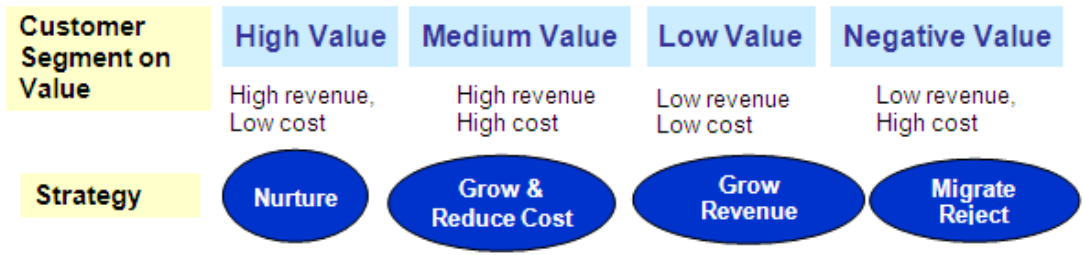

From this result, it is possible to derive some high-level business strategies. It is obvious that the best customers (considering only the data contained in the table) are falls in cluster 3. These customers have higher revenue per person than the customers of other clusters, as indicated by the total revenue column. The cluster 3 has high value, low cost and is classified as High value cluster. The cluster 1 has high revenue, high cost and is classified as Medium value. The cluster 2 has low revenue, low cost and is classified as Low value. The cluster 0 has low revenue, high cost and is classified as Negative value.

Some possible strategies include:

- A retention strategy for best customers (those in cluster 3)
- Cross sell strategy can be implanted between cluster 3 & 2 because they are close in value.
- The strategy for cluster 1 would be to wait and see. It appears to be a group of new customers for which we have not yet collected sufficient data to determine what behaviors they may exhibit.
- Cluster 0 appears to be the worst cluster, with a very low revenue percentage.

# 5. CONCLUSION AND FUTURE SCOPE

Identification of high-profit, high-value and low-risk customers via the data mining technique - customer clustering has been studied using IBM Intelligent Miner. Our organization retail smart store data is used for this study. This study uses demographic clustering technique for customer clustering. The final results demonstrate that the proposed approach revealed the high-value customers. From the results, cluster 3 has high revenue customers and represents about 35 percent of the revenue yet only by 6 percent of the customers. The cluster 3 has high value, low cost and is classified as High value cluster. The cluster 1 has high revenue, high cost and is classified as Medium value. The cluster 2 has low revenue, low cost and is classified as low value. Also, suggested possible strategies to retain and move the customers from lower band to upper band.

## 5.1 Future Scope:

Based on the results our organization retail smart store decided to undertake lot of loyalty programs for smart store customers. The further work on segmentation (clustering) using more detailed behavioral data and opportunity identification using association algorithms within the segments discovered. Other possible future works are association of products and customer segmentations for cross-selling (selling new products) and up-selling (selling more of what customers currently buy).

# ACKNOWLEDGEMENTS

Author would like to thanks to his colleague Mr. Nagarajan Karuppiah, Tata Consultancy Services for his valuable support and help.

## Author:

**Sankar Rajagopal** received M.Sc., degree in Electronics, M.E degree in Materials Science and Ph.D in Metallurgical Engineering from Bharathidasan University in 1990, 1992 and Indian Institute of Technology-Madras, Chennai 1999 respectively. After completion of his doctoral degree, he joined in TATA Consultancy Services as a Software Consultant. Then he has elevated to position Enterprise DW/BI Architect. His areas of research interests include Metal Matrix Composites, Mechanical alloying, Material Informatics, Mechanical Behaviors of Materials, Nanotechnology and Data Mining and Knowledge Discovery. He has published about 15 contributed peer reviewed papers at National / International Journals and Conferences. He received best oral presentation awards in conferences.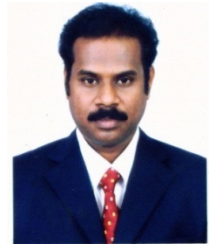